# Sparse Attention-Based Neural Networks for Code Classification

Ziyang Xiang

School of Computer Science and Technology, University of Science and Technology of China

xiangzy@mail.ustc.edu.cn

Zaixi Zhang

School of Computer Science and Technology, University of Science and Technology of China

zaixi@mail.ustc.edu.cn

Qi Liu*

School of Computer Science and Technology, University of Science and Technology of China

qiliuql@ustc.edu.cn

*Abstract*—Categorizing source codes accurately and efficiently is a challenging problem in real-world programming education platform management. In recent years, model-based approaches utilizing abstract syntax trees (ASTs) have been widely applied to code classification tasks. Many of these methods start by parsing source code into abstract syntax trees, followed by encoding them as sequences for input into temporal models or Transformer models for subsequent processing. However, existing methods suffer from issues such as loss of structural information and slow training speeds. In response to these challenges, we introduce an approach named the Sparse Attention-based neural network for Code Classification (SACC) in this paper. The approach involves two main steps: In the first step, source code undergoes syntax parsing and preprocessing. The generated abstract syntax tree is split into sequences of subtrees and then encoded using a recursive neural network to obtain a high-dimensional representation. This step simultaneously considers both the logical structure and lexical level information contained within the code. In the second step, the encoded sequences of subtrees are fed into a Transformer model that incorporates sparse attention mechanisms for the purpose of classification. This method efficiently reduces the computational cost of the self-attention mechanisms, thus improving the training speed while preserving effectiveness. Our work introduces a carefully designed sparse attention pattern that is specifically designed to meet the unique needs of code classification tasks. This design helps reduce the influence of redundant information and enhances the overall performance of the model. Finally, we also deal with problems in previous related research, which include issues like incomplete classification labels and a small dataset size. We annotated the CodeNet dataset with algorithm-related labeling categories, which contains a significantly large amount of data. Extensive comparative experimental results demonstrate the effectiveness and efficiency of SACC for the code classification tasks.

## I. INTRODUCTION

With the rise of the big data era, online intelligent education is gradually getting more attention, offering significant assistance to students in independent learning and contributing to scientific progress. Programming education, a prominent subset of online intelligent education, is gaining popularity among students with a passion for information technology. However, the extensive resources available on online programming education platforms often contain unannotated solution codes. Exploring how to add algorithmic knowledge point annotations to these codes has become a significant area of research to meet students' practical requirements. In addition, code classification serves as the foundation for various other code intelligent tasks [1][2], such as code clone detection [3][4][5] and code summarization [6][7][8]. Such research area is a classic and significant topic in the field of software engineering.

In order to classify code based on the algorithms it contains, staff members would need to read and understand code of various styles if a manual classification approach is employed. This method, which is time-consuming and requires significant effort, leads to inaccurate classification outcomes and lacks practical usefulness. In recent years, researchers have been attempting to apply methods related to natural language processing (NLP) to code classification tasks. For code analysis, a commonly used approach is the traditional text-based method, which can be mainly divided into two categories: one is the method based on word frequency [3][5], and the other is the method based on topic modeling [9][10][11]. Methods based on word frequency utilize techniques such as the Bag of Words (BoW) model [12] or the Term Frequency-Inverse Document Frequency (TF-IDF) [13] approach to calculate the frequency and weight of each word appearing in the code. Methods based on topic modeling are most commonly applied using techniques such as Latent Semantic Indexing (LSI) [14] and Latent Dirichlet Allocation (LDA) [15]. In this regard, LSI utilizes Singular Value Decomposition (SVD) [16] to reduce the dimensionality of the text and represent it as a low-dimensional vector space, followed by the application of the vector space model for text processing. LDA, on the other hand, is a method based on Probabilistic Graphical Models (PGMs), continuously updating the distribution of topics and words through iterative computation until convergence. A wealth of research indicates that these text classification methods are highly efficient and have a wide range of application scenarios. However, the limitation of traditional text-based approaches in code classification scenarios is that they treat code as plain text, ignoring its inherent nested tree-like structure, which prevents them from effectively capturing the rich and complex structural information of code.

Methods based on Abstract Syntax Tree (AST) can also be applied to code classification tasks. AST is used to represent the syntactic structure of source code. Unlike approaches that treat code as plain text, AST parsing tools can convert source code into a tree-like structure, preserving information at the logical and structural level. This method first parses the code into an AST form, then preprocesses the AST, and finally applies deep learning-related natural language processing models for subsequent processing and classification. Common approaches for handling tree-structured data include Recursive Neural Network (RvNN) [4] and Tree-based Convolutional Neural Network (TBCNN) [2]. In these, RvNN recursively computes the vector representation of each node in the tree structure, while TBCNN performs convolution operations on the tree structure to extract local features and merge them into a representation of the entire tree structure through pooling operations. After obtaining the vector representation, the AST needs to be flattened into a sequential form for subsequent processing and classification. In the field of code classification, common NLP deep learning classification models include Recurrent Neural Network (RNN) [17] and its variants, as well as Self-Attention Models [18]. RNN can process sequence data transformed from AST and perform classification [19]; Self-Attention Models can capture dependencies between different positions in the sequence [20]. Although methods based on AST have achieved significant improvements in classification accuracy compared to text-based methods [8][19][20], they suffer from long training times and the problem of vanishing gradients. This is due to the likelihood of long sequences when the AST is split and converted into a sequence in a certain way. As the sequence length increases, the RNN needs to handle more time steps, leading to an increase in model parameters and computational complexity, thereby significantly increasing training time. Self-Attention Models may also have long-term dependency problems when dealing with non-sequential input [21][22], as non-sequential structures may have long-distance dependencies. Therefore, how to effectively split and process the AST is a problem that needs to be focused on.

It can be observed that the main issues in the current research on code classification are twofold: (1) How to achieve a reasonable representation of the source code, minimizing the generated sequence length while extracting and preserving the rich semantic and structural information within the code? (2) How to select and optimize natural language classification models to better adapt to the code classification tasks with structured feature input data, thereby enhancing classification accuracy and accelerating the training speed?

To address the issues mentioned above, we propose a code classification method based on the Sparse Attention Mechanism [23], named SACC (Sparse Attention-Based Neural Networks for Code Classification). The aim is to fully mine various information within the code while improving classification accuracy and accelerating training speed. SACC reasonably utilizes the excellent performance and training speed of the model based on the self-attention mechanism, and adjusts and optimizes the self-attention mechanism by incorporating the structural information of AST for code classification tasks, thereby further enhancing the model's performance. The code classification method proposed in this paper takes code segments as input. The preprocessing part of the method first removes some redundant information from the code, then uses an AST parsing tool to parse the code segment into the form of AST. The first step is to split the AST according to the granularity of the statement. Each statement is a subtree, and this splitting method can utilize the lexical level information within the statement while preserving the overall structural information of the code. Subsequently, a recursive neural network that fits tree-structured data is used to encode the statement tree, obtaining a high-dimensional vector of the statement tree. The second step is to input the encoded statement tree sequence into the Transformer model with a sparse attention mechanism for classification. In this step, the self-attention part of the Transformer model integrates different sparse attention patterns. Among them, an AST-pattern extracted based on the code itself can provide structural information of the code for self-attention calculation, reducing unnecessary association calculations and focusing on parts that are more helpful for the classification task. Meanwhile, the sparse attention mechanism is a technique that can reduce the amount of attention computation and accelerate training. Therefore, our method can more efficiently utilize the attention mechanism, improving the accuracy and training speed of the model. Finally, we construct a collection of algorithmic knowledge points suitable for code classification by integrating multiple existing Online Judge (OJ) systems, making the classification results more practical and interpretable. Extensive experiments were conducted on a large-scale code dataset from Online Judge systems, which verified the accuracy and efficiency of SACC. In summary, the innovations of this paper are as follows:

- SACC innovatively combines abstract syntax trees and sparse attention mechanisms for code classification. It transforms the code into an AST form, carries out reasonable splitting and encoding, integrates the structure of the AST into the self-attention computation, and utilizes lexical and structural level information, thereby improving the accuracy of classification.

- The application of the sparse attention mechanism to classification tasks reduces the computation of self-attention, significantly improving computational speed and making the model efficient.

- Aimed at practical application scenarios, a collection of algorithmic knowledge points for code classification has been constructed. Extensive experiments have been conducted to verify the effectiveness of our proposed method SACC.

The rest of this paper is organized as follows: Section 2 of this paper introduces the related works of code representation and code classification; Section 3 presents the problem definition and preliminary knowledge; Section 4 elaborates on the basic framework and technical methods of SACC; Section 5 describes a multitude of experiments conducted around SACC, verifying its effectiveness and efficiency; Finally, Section 6 concludes the paper and proposes future prospects.

## II. RELATED WORK

In this section, we will introduce the related work on code algorithm classification from two aspects: code representation and sparse attention.

### A. Code Representation

In the field of software engineering, code serves as a crucial form of data, and the effective representation of code is a fundamental research question [24].

Some traditional code representation methods treat code as plain text and perform feature extraction at the lexical level. Representation methods based on word frequency consider the frequency of each word's occurrence in the text as a feature. For instance, SourcererCC [3] is a code clone detector based on the BoW model, representing identifiers and keywords in code segments using word frequencies. It employs an optimized inverted index to rapidly query potential clone relationships for given code pairs. Frantzeskou et al. [1] addressed the problem of source code authorship classification by employing N-gram language models to represent code. They used various machine learning algorithms like decision trees, support vector machines, and naive Bayes for code classification. Reference [25] treated code as documents, employing the TF-IDF method to calculate word frequencies in the code and performed programming language classification by combining multi-source information. Approaches based on topic modeling consider code as text composed of topics, with each topic consisting of specific words. The objective of this method is to learn these topics and the distribution of each topic within code texts. Maletic et al. [26] used Latent Semantic Indexing (LSI) from information retrieval to compute semantic similarity between software components, addressing program comprehension tasks in software system maintenance and refactoring. In order to improve the accuracy of software fault prediction, the MWE [6] approach employed LDA technology and information entropy measurement to assess the cohesiveness of different classes within software.

In certain studies within the field of software engineering, researchers have employed graphical data structures that involve extracting control flow and data dependency relationships from code. They have used relevant methods of graph embedding [27] for code representation analysis. A Control Flow Graph (CFG) is a graphical representation that depicts program statements and control structures as nodes, with directional edges representing control flow relationships between nodes. Tufano et al. [28] extracted CFGs from code to predict code similarity using various representation forms, while other efforts [29] combined control flow and data flow within a Program Dependence Graph (PDG) for representation. However, such representation methods heavily rely on intermediate representations generated from parsed code. In practical scenarios, incomplete or non-compilable code fragments are common, making graph-based code representation methods hard to apply.

The abstract syntax tree offers a language-independent intermediate representation that makes code easier to understand and analyze due to its structured form. In recent years, numerous researchers have engaged in research on AST-based representations. Once code is parsed into an AST, the key focus of representation lies in effectively mining lexical and structural level semantic information. TBCNN [2] utilizes a customized tree-based neural convolutional network to learn vector representations of code on the AST. Binary tree structures are employed to address the variable number of child nodes in AST, with designed tree convolutional kernels capturing structural information. In order to avoid potential information loss, researchers have initiated research into representation methods that preserve the original structure of AST. DeepCom [30] automatically generates developer comments for Java methods, using a Structural-Based Traverse (SBT) algorithm to flatten the AST into a sequence while enclosing subtrees under the same node with parentheses to retain the original AST structure and prevent information loss. CDLH [31] employs Tree-LSTM for AST representation, recursively computing vector representations for each node during traversal. For larger ASTs that lead to long sequences after parsing, ASTNN [19] splits the AST at the statement level and reduces the sequence length. It then employs Bidirectional Gated Recurrent Units (Bi-GRU) to capture dependency and contextual relationships within the sequence. To restore the original structure of AST, CAST [8] applies RvNN to reaggregate and update the vector representations of subtrees according to the tree relationship before splitting. These representation methods focus more on the contextual information of AST sequences and have performed well in various tasks. Therefore, our work adopts the AST-based code representation approach.

### B. Sparse Attention

The self-attention mechanism, introduced by Vaswani et al. [18] in the study of the Transformer model, allows learning dependencies between different positions in a sequence, enabling a better capture of contextual information in the sequence. Initially used for machine translation in the field of natural language processing [18], the Transformer has subsequently found widespread applications in various domains including computer vision [32], speech recognition [33], and recommendation systems [48].

The Transformer model and the self-attention mechanism have had a huge impact on the development NLP. Prior to the introduction of the Transformer, sequence models such as RNNs [17] and LSTMs [34] were commonly used for various NLP tasks. However, these sequence models faced challenges in handling long sequences due to issues like vanishing or exploding gradients. Additionally, their sequential computation resulted in slower training speeds. The Transformer model utilizes the self-attention mechanism to capture dependencies between any two positions in a sequence. It computes attention scores for all positions simultaneously, avoiding long-term dependency problems and improving both model performance and training speed. The Transformer has also been applied to code summarization tasks [35], with adjustments made to the attention mechanism. However, in contrast to the sequential nature of natural language, the flattened sequences obtained from ASTs possess multi-level nested structural features. Researchers have pointed out [22] that the Transformer can still face long-term dependency issues when dealing with long sequences containing nested structures.

The self-attention mechanism, when dealing with long sequences, requires the computation of attention scores between all positions, resulting in quadratic-level temporal and spatial complexity in most tasks. This leads to slower training and inference speeds for the model. Sparse attention is a technique introduced by Child et al. [23] to reduce the computational cost of the self-attention mechanism. It involves setting a portion of positions in the self-attention mechanism to zero, thus enhancing the model's efficiency. Both Longformer

[36] and BigBird [37] have explored various sparse attention mechanisms to improve the performance of Transformers when dealing with long sequences. Reformer [45] devised the use of Locality Sensitive Hashing (LSH) to group similar attentions together, approximating the final scores. Linformer [46] experimentally demonstrated that the attention matrix in Transformers is of low rank. They performed Singular Value Decomposition (SVD) on it and modified the self-attention structure to achieve linear complexity in computation. The mentioned efforts have been primarily focused on the domain of NLP. Currently, the focus of research is on designing appropriate sparse attention patterns suitable for specific tasks to improve model performance.

## III. BACKGROUND

In relation to the conducted work, this section introduces the problem definition and relevant knowledge of the SACC code classification method.

### A. Problem Definition

In the context of code algorithm classification for programming education, given a collection of code snippets $C = \{C_1, C_2, ..., C_S\}$, and a predefined set of code algorithmic knowledge point labels $L = \{L_1, L_2, ..., L_P\}$, each code snippet $C_i$ is associated with a knowledge point from the set $L$.

The research objective is to establish a model capable of accurately classifying given code snippet samples into corresponding knowledge point labels. To achieve this goal, a function $f(C)$ can be used: a mapping from the set of code snippet samples $C$ to the set of knowledge point labels $L$, where each input code snippet is mapped to an element within the knowledge point label set. To fit this function $f(C)$, a parameterized model is employed, with parameters $\theta$ that can be learned and optimized through a training process. Thus, our task is to find the optimal parameters $\theta^*$ such that, for a given code snippet sample $C_i$, the function $f(C_i; \theta^*)$ can accurately predict its corresponding knowledge point label $L_i$. Therefore, the problem of code algorithm classification for programming education can be defined as follows.

***Definition*** Given a set of code segments $C$ and a set of knowledge points $L$, where each code segment is associated with an algorithmic knowledge point in $L$. The main research objective of this paper is to optimize the parameters $\theta$ of the model, fit a function $f(C)$ that can correctly classify the code segments to their corresponding knowledge point labels, thereby improving the performance of the model on the code classification task.

### B. Abstrct Syntax Tree

Due to the inherent tree-structured features of the code, the AST is a data structure widely used in software engineering and code analysis tasks. Compared to the original data, AST omits details irrelevant to code representation, such as punctuation, spaces, and indentation symbols, making it a more abstract representation focusing on code semantics and structure. After lexical and syntax analysis, a piece of C language code as shown in Figure 1(a) can be parsed into the AST in Figure 1(b), where some of the specific values contained in the nodes are shown. In an AST, starting from the root node representing the entire code snippet, it connects to multiple leaf nodes representing the smallest syntax units, with the intermediate nodes representing syntax structures. Each node, except for the root node, has and only has one parent node, and each node can have multiple child nodes. To systematically read the information contained in the AST, we can convert the AST into a token sequence using preorder traversal. Some studies have pointed out that simple traversal can lead to the loss of structural information contained in the AST [19][31]. Moreover, due to the usually large number of nodes and depth of the tree contained in the AST, direct traversal will result in a very long sequence, which is not beneficial to subsequent processing. Therefore, it is a common technique to first reasonably split the AST and construct a shorter subtree sequence.

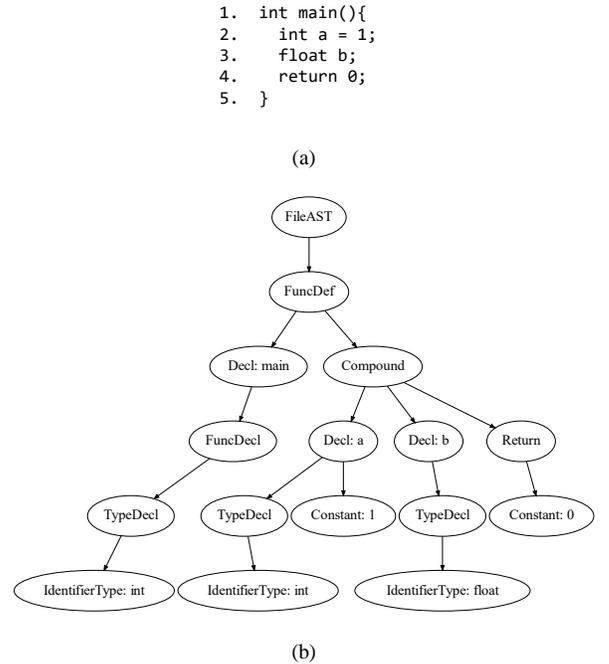

Figure 1. A code sample and its corresponding AST. Some specific values are shown after the tokens of the nodes.

### C. Transformer

This section introduces the self-attention mechanism and the related knowledge of the Transformer model proposed by Vaswani et al. [18]. In NLP tasks, traditional methods are mostly based on recurrent neural network sequence models, while the Transformer eliminates the dependence of sequence models on time steps and calculates the correlation between different positions in the sequence in parallel. Figure 2 shows the basic structure of the Transformer encoder, where the most core part is the self-attention module. As can be seen, the model first positionally encodes the input sequence vector to retain the sequential information; then inputs the sequence into $N$ connected Transformer encoders, each of which has the same structure and parameters, and can be divided into two sub-layers: Multi-head Self-Attention Layer and Feed-Forward Neural Network Layer. After each sub-layer, residual connections and layer normalization techniques are also applied to improve the performance and efficiency of network training.

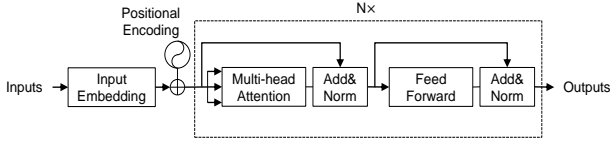

Figure 2. Basic structure of Transformer encoder. The model takes inputs with positional encoding and feeds it into the encoding layer with two sublayers. The output is obtained by passing through multiple encoding layers.

*1) Self-attention Module*

The design inspiration for the attention mechanism comes from the study of human vision. When observing objects, humans focus their attention on the target area that needs to be focused on and ignore irrelevant information. Bahdanau et al. [38] first introduced the attention mechanism in the task of neural machine translation. During the decoding process, the model can dynamically focus its attention on different positions of the input sequence and selectively pay attention to the input content related to the current output. Specifically, let $X = [X_1, X_2, ..., X_N]$ be the high-dimensional input vector sequence after encoding, given a query vector $q$, the probability $\alpha_i$ that the model chooses to pay attention to the i-th input element can be calculated as follows:

$$\alpha_i = softmax(f(x_i, q)) = \frac{\exp(f(x_i, q))}{\sum_{j=1}^{N}\exp(f(x_j, q))} \quad (1)$$

$f(x_i, q)$ is the attention score calculation function for the i-th element $x_i$ in $X$ and $q$, which can be roughly divided into additive attention [38] and multiplicative attention [39].

The attention mechanism associates and weights different sequences, while the self-attention mechanism focuses on the dependencies between different positions within the sequence. This method first obtains the query, key, and value from the input sequence $X$, mapping them to the query vector $Q$, key vector $K$, and value vector $V$; it then calculates the attention score between $Q$ and $K$ to obtain the relevance of different positions in the sequence; then, a softmax function is used to normalize the attention scores to obtain attention weights; finally, the attention weights are multiplied with $V$ to obtain the weighted representation. The entire process can be represented by (2):

$$Attention(Q, K, V) = softmax(\frac{QK^T}{\sqrt{d_k}})V \quad (2)$$

The scaled dot-product attention [18] is used to calculate the attention score between $Q$ and $K$, where $d_k$ is the dimension of the $Q$ vector and $K$ vector.

Sequential models like RNNs often see a gradual decay in information transmission as the number of time steps increases. The computation of self-attention, however, can capture the dependency between elements at any two positions in the sequence, possessing the capability to model long-range dependencies; on the other hand, the self-attention module can utilize matrix multiplication techniques to parallelize the computation of attention between elements in the sequence, significantly accelerating computation speed compared to sequential models.

*2) Multi-head Attention Mechanism*

To capture a more comprehensive range of dependencies within the sequence, the multi-head attention mechanism applies the self-attention mechanism to multiple different projected vector spaces. For each set of projected vectors, the calculation is performed according to (2), and then the outputs of these attention heads are concatenated to obtain the final representation. The calculation process is as follows:

$$MH(Q, K, V) = Concat(H_1, ..., H_h)W^O \quad (3)$$

$$H_i = Attention(XW_i^Q, XW_i^K, XW_i^V) \quad (4)$$

Where $W_i^Q \in \mathbb{R}^{d_{model} \times d_k}$, $W_i^K \in \mathbb{R}^{d_{model} \times d_k}$, $W_i^V \in \mathbb{R}^{d_{model} \times d_k}$, $W^O \in \mathbb{R}^{hd_k \times d_{model}}$. Here, $h$ represents the number of attention heads and $d_{model}$ represents the dimension of the input sequence $X$. The design of multi-head attention still allows for efficient parallel computation and enhances the model's generalization ability.

*3) Positional Encoding:*

Although the design of the self-attention mechanism can calculate the correlation between elements at any two positions in the sequence, the sequential information is not reflected in the model. To supplement this, the Transformer adds positional encoding before it is input into the attention layer. The encoding obtained through (5) and (6) carries a position-related texture information, which preserves the characteristics of natural language.

$$PE_{(pos, 2i)} = \sin(pos / 10000^{2i/d_{model}}) \quad (5)$$

$$PE_{(pos, 2i+1)} = \cos(pos / 10000^{2i/d_{model}}) \quad (6)$$

*4) Feed-forward Network*

After the output results of the multi-head attention layer, a feed-forward neural network is used to extract and map the feature information therein, providing a more rich and accurate representation for subsequent tasks. As shown in (7), the feed-forward neural network consists of two fully connected linear layers:

$$FFN(x) = \max(0, xW_1 + b_1)W_2 + b_2 \qquad (7)$$

The first fully connected layer maps the input vector to a higher dimension using a ReLU activation, and the second layer maps the high-dimensional features back to the original dimension.

*D. Sparse Attention*

The Transformer has achieved significant results in multiple fields such as NLP [18], computer vision [32], recommendation systems [48], and protein structure [40]. However, under the dot-product attention calculation method, the time and space complexity of self-attention will increase quadratically with the length of the sequence. Sparse attention mechanism is a variant of the attention mechanism, aiming to reduce the computational and storage overhead. Traditional attention computation suffers from performance degradation and excessive computation time when dealing with longer sequences. Sparse attention only calculates a portion of the key values related to the query to reduce the impact of redundant information on the model effect and improve computational efficiency.

In the design of sparse attention, the most critical part is how to select the keys and values that need to be computed according to task requirements. Child et al. [23] combined local self-attention and dilated self-attention in the research of sparse attention, learning both local and remote sparse correlations. On this basis, Longformer [36] added a global attention sparse pattern to capture the correlation between some key positions in the sequence and all other positions. BigBird [37] added a random attention pattern.

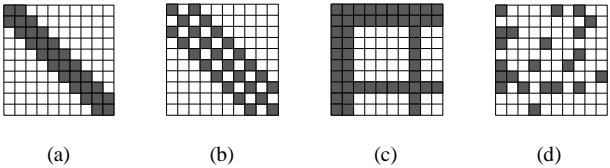

(a)      (b)      (c)      (d)

Figure 3. Sparse attention patterns. (a) shows the local pattern with window size 3, (b) depicts the dilated pattern similar to dilated convolution, (c) is the global pattern with global size 3 and (d) represents a randomly generated random pattern.

Figure 3(a) shows the local attention pattern, also known as the sliding window pattern, which indicates that each element in the sequence only calculates the attention score with $w$ elements around itself, including itself, where the window size $w$ in the figure is 3. Figure 3(b) is the dilated attention pattern, inspired by dilated convolution in computer vision [41]. Figure 3(c) is the global attention pattern with the number of global elements $g = 3$ in the figure. Figure 3(d) is the randomly generated attention pattern. These patterns mainly target tasks in natural language processing, capturing different types of key information in text sequences well. However, when facing code classification tasks, these attention patterns may still cause the loss of key information, such as the tree structure of the code.

## IV. MODEL ARCHITECTURE

To tackle the problems encountered by plain text representation and existing decoding models, we propose a new code classification method based on sparse attention mechanism, SACC. This method fully utilizes the semantic richness of the abstract syntax tree representation, combines the efficiency of Transformer and the information conciseness of sparse attention, making the final classification result highly accurate.

*A. Overall Architecture*

This section presents the overall architecture of SACC proposed in this paper. The entire process is depicted in Figure 4. This method can be summarized as follows:

- Tools such as pycparser are used to parse the input source code syntactically, obtaining the data form of AST. Then it is traversed and split to extract AST attention pattern based on the structure. The statement subtrees are input into the RvNN to extract lexical information and internal structures. The subtrees are encoded into a high-dimensional vector to obtain the sequence of the statement trees.

- The AST pattern from the previous step is integrated with other sparse attention patterns. The statement tree sequence is input into a Transformer encoder with a sparse attention mechanism for feature extraction. After passing through the pooling layer, the classification result of the code is predicted.

The model details will be elaborated in the following sections.

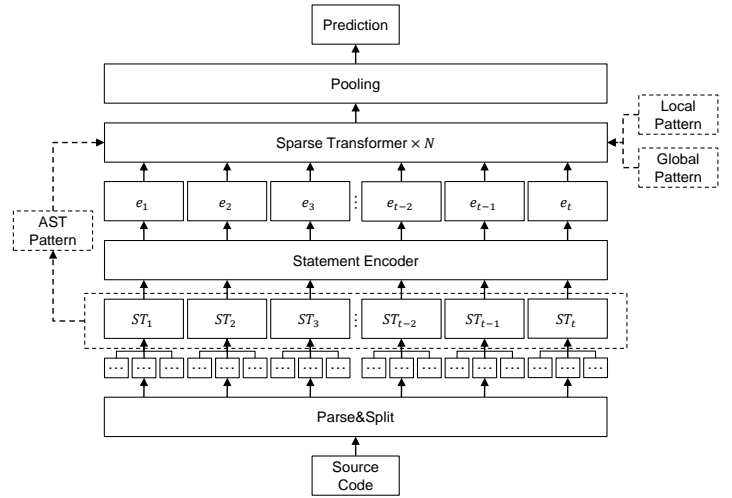

Figure 4. The overall framework of SACC. The SACC architecture takes a given source code as input, extracts the AST pattern during encoding, and integrates various sparse attention patterns into the Transformer. It then produces classification predictions for the given code.

## B. Specific Steps

As mentioned above, existing methods have issues such as incomplete capture of structural information and longer model training time. Therefore, we propose SACC to more comprehensively obtain information in the code's AST and accelerate the training process.

### 1) Parsing and Preprocessing

If we treat the source code as plain text, there are usually many redundant pieces of information, such as spaces, indents, and punctuation required by syntax, etc. Moreover, due to different writing styles, authors may write different codes even when expressing the same semantics. To remove irrelevant characters and obtain a unified semantic expression, we first use the parsing tool pycparser to parse the source code into an AST. Figure 5 shows a code snippet from the experimental data and its parsed structure. For code that is not particularly short, the AST usually has many nodes and a large depth. If the entire AST is used as the subsequent input, there would be a problem of gradient vanishing when encoding it with RvNN. If the AST is split at the node level, the representation will degenerate into a BoW model, losing the structural information of the code. Therefore, we use the granularity of statements to split the AST, achieving a balanced trade-off between subtree size and the richness of structural information, as shown in Figure 5. Specifically, each common statement is split as a statement tree. When the statement is of the *FuncDecl*, *If*, *While*, and other types, these statements will be followed by a block. The part before the block is considered part of this statement, and the body of the block continues to be split according to this rule. As a result, the AST of a code snippet becomes a sequence composed of statement trees.

We notice that the statement trees formed by this splitting rule also have tree-like relationships with each other. During splitting, we can construct such relationships and extract the AST sparse attention pattern corresponding to the code, as shown in Figure 6. Specifically, we will obtain the adjacency matrix corresponding to this tree-like relationship. Elements with a value of 1 represent the need to calculate the attention score between these two statement tree elements during attention computation, while those with a value of 0 will not be calculated.

### 2) Encoding of Statement Tree

After obtaining the statement tree sequence, in order to fully extract the lexical and semantic information within each subtree, an RvNN with a tree structure is used to encode it. Specifically, word embedding training is first conducted on the corpus composed of tokens from all AST nodes [42], converting each node's corresponding word into a high-dimensional representation to capture the semantic information between vocabularies. This pre-training yields the matrix $W_e^T \in \mathbb{R}^{|V| \times d}$, where $|V|$ is the size of the node token corpus and $d$ is the dimension of the word embedding. Given a statement tree, its node $n$ can be represented as:

$$w_n = W_e^T x_n \tag{8}$$

Where $x_n$ is the one-hot encoding representation of node $n$, which is transformed into a high-dimensional vector $w_n$ after word embedding. Then, RvNN is used to encode all nodes within the statement tree:

$$h = \sigma(W_n^T w_n + \sum_{i=1}^{C} h_i + b_n) \tag{9}$$

Where $h$ is the hidden state, $\sigma$ is the activation function, $W_n^T \in \mathbb{R}^{d \times k}$ is the encoding matrix, $k$ is the dimension after encoding, $C$ is the number of child nodes of node $n$, and $b_n$ is the bias term. In this way, the vector representations of all nodes within the statement tree can be computed recursively, and the vector representation of the entire statement tree is obtained at the end using max pooling.

### 3) Sparse Attention Transformer Classification

After encoding the statement tree, we can obtain the statement tree sequence $E = \{e_1, e_2, ..., e_N\}$. In order to maintain the sequential nature, position encoding is required before the attention module, as shown in Figure 4.

Firstly, SACC adopts a sliding window pattern. It is described using symbolic language as follows:

$$A_{ij} = \begin{cases} Q_i K_j^T, & |i-j| < \lfloor w/2 \rfloor \\ 0, & otherwise \end{cases} \tag{10}$$

Where $A_{ij}$ is the attention score between $e_i$ and $e_j$ in the statement tree sequence $E$, $Q$ and $K$ are the query and key matrices respectively, and $w$ is the size of the sliding window. As shown in Figure 3(a), the sliding window pattern aims to calculate the association between each statement tree and the others within a distance of no more than $\lfloor w/2 \rfloor$ from itself, thereby capturing the local correlation in the sequence.

In some text-related tasks [23], it is necessary to capture long-distance dependencies. To retain the information contained in some elements that play a significant role in the overall semantics, SACC also employs a global attention pattern as shown in Figure 3(c):

$$A_{ij} = \begin{cases} Q_i K_j^T, & i \in G \ or \ j \in G \\ 0, & otherwise \end{cases} \tag{11}$$

Where $G$ represents the elements set by the global attention pattern.

The above attention patterns are designed for NLP tasks, which cannot capture structural information when dealing with code. Figure 6 shows the construction method of the AST attention pattern. SACC captures the tree structure relationship between statement trees when splitting the original AST and stores it as an adjacency matrix $A_{ij}$. The attention pattern based on AST can be expressed as:

$$A_{ij} = \begin{cases} Q_i K_j^T, & Adj[i][j]=1 \\ 0, & Adj[i][j]=0 \end{cases} \quad (12)$$

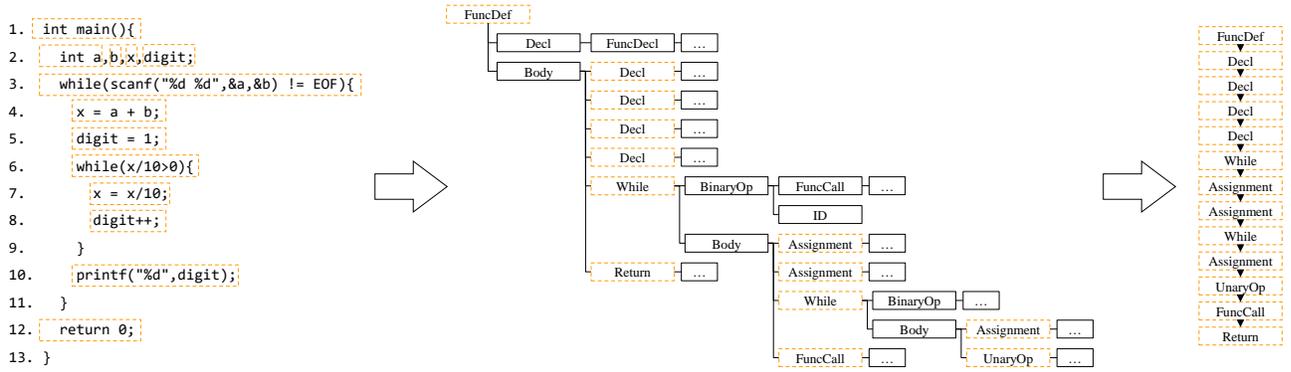

Figure 5. Parsing and splitting of the AST. The statement nodes and their corresponding code lines are highlighted in the figure after parsing the code sample into an AST, the statement nodes and their corresponding code lines are highlighted in the figure. After splitting, a sequence of statement trees can be obtained.

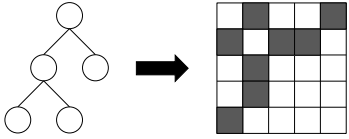

Figure 6. AST sparse attention pattern. The tree is stored in in the form of an adjacency matrix, which represents the AST pattern.

Thus, the sparse attention can be integrated into (13), i.e., if the sequence pair $(i, j)$ belongs to the above patterns, the attention score between $e_i$ and $e_j$ is calculated, otherwise, it is directly set to 0.

$$A_{ij} = \begin{cases} Q_i K_j^T, & (i,j) \in predefined\ patterns \\ 0, & otherwise \end{cases} \quad (13)$$

The rest of the sparse attention Transformer remains consistent with the original design [18]. After the input is passed through the encoder structure, the extracted features are sent to a pooling layer to obtain the final classification result.

### C. Model Learning

SACC predicts the algorithmic knowledge points contained in the code, which is a typical classification task. Given the number of classification categories $P$, the code vector $v$, a matrix can be obtained through SACC to project $v$ onto a space with a dimension of $P$, thereby obtaining a probability vector of different classifications $\hat{x} = W_o^T v + b_o$, where $W_o \in \mathbb{R}^{d \times P}$, $d$ is the dimension of the code vector, $b_o$ is the bias term, and the cross-entropy loss function is used:

$$J(\Theta, \hat{x}, y) = -\sum (\log \frac{\exp(\hat{x}_y)}{\sum_j^P \exp(\hat{x}_j)}) \quad (14)$$

Where $\Theta$ represents all the parameters of the weight matrices in SACC, and $y$ is the ground truth.

## V. EXPERIMENTS

### A. Dataset Description

The dataset used in the experiment is from the CodeNet project [43], with data sourced from two Japanese OJ systems, AIZU and AtCoder. The project inherits the concept of the ImageNet [44] dataset, containing nearly 14 million code samples, approximately 500 million lines of code, and 55 different programming languages, making it a large-scale, diverse, and high-quality dataset. In addition, the dataset also includes content descriptions of the problems and the submission results from users.

As the original data did not include any labels, we constructed an algorithmic knowledge point collection with 18 types based on the categories of multiple OJ systems. They are: (1) Brute Force, (2) Greedy, (3) Dynamic Programming, (4) Recursion, (5) Search, (6) Sorting, (7) Simulation, (8) Arithmetic, (9) Combination, (10) Game Theory, (11) Data Structure, (12) Graph Theory, (13) Computational Geometry, (14) String, (15) Bit Manipulation, (16) Discretization, (17) Random Algorithm, (18) NP Problem. Combining the descriptions of programming problems in the dataset with the above algorithmic knowledge points, all problems were manually labeled. Code in the dataset that compiles and achieves an "AC" (Accepted) execution result is considered to meet the problem requirements, i.e., the code corresponds to the knowledge point involved in its associated programming problem.

In this study, we selected the C language as the language of choice for this classification task. After suitably processing the source code, we obtained 196,662 code samples that could be parsed, indicating a significant dataset size increase, which is nearly four times larger than the 52,000 code samples used in previous research with the poj104 dataset. Given that the majority of the annotated results in the code are single-label classifications, this task can be regarded as a single-label classification task. After preprocessing the code in a way that does not affect its semantics, we utilized pycparser to parse it into an AST form. The statistical data for the final dataset is presented in Table 1. Figure 7 provides a label statistical overview of the data.

TABLE I. DATASET STATISTICS

| Statistical information | |
|---|---|
| #Programs | 196662 |
| #Classes | 18 |
| Max AST nodes | 100354 |
| Avg. AST nodes | 133.22 |
| Max AST depth | 398 |
| Avg. AST depth | 11.01 |

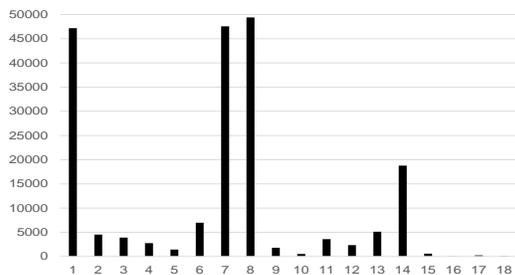

Figure 7. Count of different categories in the dataset.

### B. Evaluation Metrics

This experiment adopts Accuracy, Precision, Recall, and F1-score as the evaluation metrics. These metrics are widely used in assessing classification models, and their definitions are as follows:

$$Accuracy = \frac{TP + TN}{TP + FP + FN + TN}$$

$$Precision = \frac{TP}{TP + FP}$$

$$Recall = \frac{TP}{TP + FN}$$

$$F1 = \frac{2 \times P \times R}{P + R}$$

Where $TP$, $TN$, $FP$, and $FN$ represent the samples predicted as True Positive, True Negative, False Positive, and False Negative, respectively. Accuracy represents the ratio of the number of samples correctly predicted to the total number of samples; Precision refers to the ratio of samples predicted as positive to the actual positive samples; Recall is the proportion of actual positive samples correctly predicted as positive; F1-score is the harmonic mean of Precision and Recall, balancing the model's classification accuracy and coverage. To address the numerical differences between samples of different categories, we adopt the calculation method of macro averages, i.e., calculating the above indicators for each classification category separately, then taking the average as the final result.

### C. Baseline Approaches

To validate the effectiveness of SACC, we compare SACC with the following baselines:

- Plain text + LSTM [34]. The code is treated as plain text, tokenized according to identifiers, operators, and separators, etc. The sequence is then embedded into LSTM to extract semantic information for classification. The memory cell structure of LSTM enables it to effectively handle long-term dependencies, and it has been widely used in the field of natural language processing.

- ASTNN [19]. Using Bi-GRU, a variant of LSTM, to extract features from the statement tree sequence. This model can consider the past and future context information in the input sequence simultaneously, making it an excellent tool for handling time-series data.

- TBCC [20]. This method uses Transformer to process the statement tree sequence. It utilizes the Transformer's ability to capture the associations between any different positions in the sequence, thereby more fully extracting the semantic information in the code. To our knowledge, TBCC is currently the state-of-the-art method for code classification tasks.

### D. Experimental Results

The experiments in this section will be conducted from the following three aspects:

- Through comparative experiments, we explore whether SACC has a performance improvement in classification tasks compared to other baselines.

- Perform ablation experiments under different sparse attention patterns to explore the degree of impact of different patterns on model performance improvement.

- By comparing the model training speed, we explore whether the sparse attention mechanism of SACC can improve computational efficiency.

In terms of experimental settings, the number of training epochs is 30, the batch size is 16, the encoding dimension $d$ is 128, the number of attention layers is 2, the dimension of query key value encoding $d_k$ and $d_v$ are 64, the feed-forward layer dimension $d_{ff}$ is 2048, the sliding window size $w$ is 3, the Adamax optimizer [47] is used, and the learning rate is 0.002. All experiments are conducted on a CPU with 32 cores at 2.30GHz and a Tesla V100 GPU.

*1) Code Classification Performance*

TABLE II. PERFORMANCE COMPARISION

| Baseline | Accuracy (%) | P | R | F1 |
|---|---|---|---|---|
| Plain Text + LSTM | 82.88 | 62.17 | 58.26 | 59.99 |
| ASTNN | 89.44 | 78.86 | 70.76 | 72.67 |
| TBCC | 91.89 | 81.55 | 75.48 | 76.29 |
| SACC | **92.35** | **83.28** | **76.09** | **77.51** |

As shown in Table 2, the accuracy of the ASTNN method is 6.56 percentage points higher than that of the plain text + LSTM, and the other metrics are also significantly superior. This is because treating code as plain text will lose rich semantic structure information, and the sequence length of plain text is too large, which will still cause information loss during propagation; if truncation is used to limit sequence length, it will directly cause a large amount of valid information to be missing. The representation based on AST can better extract the tree structure features of the code, and the sequence length can be effectively controlled by splitting the AST.

The performance of TBCC have improved compared to ASTNN. This is because the sequence of code statement trees does not inherently possess strong temporality, and the Transformer can capture the associated information over longer distances in the sequence. It can also increase the model capacity to handle more complex information by stacking layers.

SACC outperforms TBCC on all metrics. This is because the Transformer also faces long-term dependency when processing sequences with complex structural information. The sparse attention pattern designed by SACC allows the model to pay more attention to the structural information of the AST and some other important information, reducing the impact of redundant information in the sequence on the model, thereby enhancing the effectiveness of the model.

*2) Ablation Studies*

TABLE III. ABLATION STUDIES

| Method | Accuracy (%) | P | R | F1 |
|---|---|---|---|---|
| SACC | **92.35** | **83.28** | **76.09** | **77.51** |
| w/o local pattern | 91.84 | 77.50 | 75.82 | 76.21 |
| w/o global pattern | 91.80 | 78.59 | 75.27 | 76.15 |
| w/o AST pattern | 91.57 | 78.55 | 75.82 | 76.45 |

The results of ablation experiments on different sparse attention patterns are shown in Table 3. It can be seen that the performances decrease to varying degrees under the setting of reducing a specific attention pattern. This is because the local attention pattern can capture the associated information between the statement tree and its surrounding statement trees; the global attention pattern can set a global statement tree, preserving its attention calculation with all other statement trees; and the AST sparse attention pattern can make the model focus on the tree structure inherent in the code, which is a key design for code classification. We noticed that the experimental accuracy decreased by 0.78 percentage points when the AST sparse attention pattern was removed, which had the greatest impact on the experimental results in the ablation experiment. This fully illustrates the importance of AST's structural information in this task.

*3) Training Speed*

TABLE IV. COMPARISON OF TRAINING SPEED

| Baseline | Training Time Per Epoch (s) |
|---|---|
| ASTNN | 1485.53 |
| TBCC | 829.16 |
| SACC | **747.29** |

As shown in Table 4, the comparison experiment of training speed only involves ASTNN, TBCC, and SACC, as the method of plain text + LSTM does not include the part of AST encoding. Under the same total number of epochs, the average training time of a single epoch for TBCC is significantly shorter than that of ASTNN. This is because the LSTM-based sequential model is a type of recurrent neural network that needs to process the elements in the sequence one by one according to the input order. Due to this structural feature, the computation of LSTM at each time step depends on the output of the previous time step. However, TBCC uses Transformer for the processing part of the statement tree sequence. Its self-attention mechanism can consider all positions in the sequence at the same time, enabling the Transformer to perform highly parallel computations, thereby improving the training speed of the model.

The average training speed of SACC is faster than TBCC. This is because in traditional full attention calculation, each query has to be associated with all positions in the sequence, leading to a quadratic increase in computational complexity with the length of the sequence. However, the sparse attention mechanism restricts the scope of attention calculation, allowing each query to be associated with only a small part of the key positions, reducing computational complexity. Therefore, it speeds up the training of the model when dealing with longer sequences.

*E. Analysis*

From the experimental results of the three parts, it can be seen that the code classification method SACC proposed in this paper, which is based on the sparse attention mechanism, has a good effect in classifying the algorithmic knowledge points contained in the code. This is because this method considers both the lexical information and tree structure information of the code, designs a sparse pattern in the attention calculation part, reduces the impact of redundant information on the model, and improves the calculation speed.

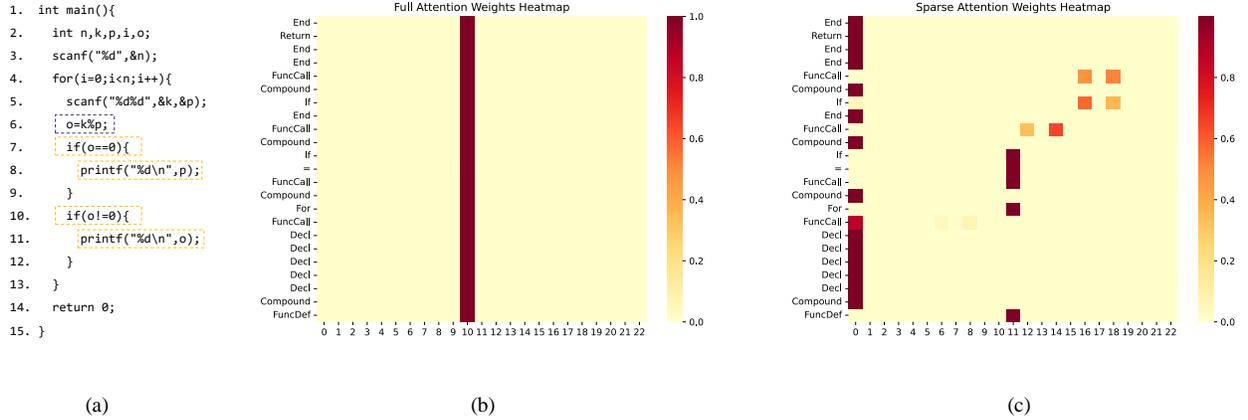

Figure 8. Attention Visualization. The parts helpful for code classification are highlighted in (a). The sequence of statement trees obtained after splitting the code sample is displayed in (b) and (c) coordinates. The heatmaps represent full attention and sparse attention, respectively.

In order to investigate the role of the sparse attention pattern of SACC on the model, we select two methods, the full attention pattern and the sparse attention pattern, to visualize the self-attention weight matrix, as shown in Figure 8.

The visualization selected a code sample with the classification label of Arithmetic. After parsing and splitting, the statement tree sequence shown in the coordinates of Figure 8(b) and 8(c) can be obtained, where the sequence number of the horizontal axis corresponds one-to-one with the name of the statement tree on the vertical axis. By reading and observing, it can be known that the crucial modulus operation is performed in the 6th line of the code, and a simple arithmetic logic is implemented in combination with the subsequent *if* judgment statement. The horizontal coordinate of the statement tree corresponding to the 6th line is 11, and the sequence numbers of the statement tree for the conditional judgment are 12, 14, 16, and 18 respectively. As shown in Figure 8(b)(c), due to the lack of restrictions on the attention calculations in the full attention pattern, sometimes all attention will be focused on invalid positions, which affects the prediction; whereas the sparse attention pattern designed by SACC concentrates more attention on the modulus operation statement tree, and under the regulation of the AST attention pattern, some attention is focused on the subsequent conditional judgment structure, which is beneficial for the model to extract comprehensive information to make correct judgments.

## VI. CONCLUSIONS

In the domain of code classification tasks, conventional approaches like text classification, sequential models, and Transformers exhibit certain limitations, namely, they disregard the structural information inherent in the code, and handling long sequences is especially time-consuming. We propose a code classification algorithm SACC based on the sparse attention mechanism. The input of the model is a piece of source code; SACC first parses the source code into an abstract syntax tree, retaining the logical and lexical level information, and preprocesses it to obtain the statement tree sequence and AST sparse attention pattern for subsequent processing; then, the statement tree sequence is input into the recursive neural network to encode into a high-dimensional representation form, fully utilizing the tree-structured data characteristics of the statement tree; then, the encoded statement tree sequence is input into the Transformer model that integrates different sparse attention patterns for classification, combining the efficiency of the attention mechanism and the structural information of AST. Finally, extensive experiments prove the accuracy and efficiency of SACC. The method proposed in this paper is only carried out on the task of code algorithmic knowledge point classification, and the needs in actual application scenarios may be more diversified, such as code difficulty prediction, code multi-label classification and other tasks. How to optimize and expand our model to cope with more challenges is one of the future research directions.

ACKNOWLEDGMENT

This research was partially supported by a grant from the National Key Research and Development Program of China (Grant No. 2021YFF0901003).

REFERENCES

[1] Frantzeskou, G., MacDonell, S., Stamatatos, E., & Gritzalis, S. (2008) Examining the significance of high-level programming features in source code author classification. Journal of Systems and Software, 81(3), 447-460.

[2] Mou, L., Li, G., Zhang, L., Wang, T., & Jin, Z. (2016) Convolutional neural networks over tree structures for programming language processing. In: Proceedings of the AAAI conference on artificial intelligence. Phoenix. Vol. 30. No. 1.

[3] Sajnani, H., Saini, V., Svajlenko, J., Roy, C. K., & Lopes, C. V. (2016) Sourcerercc: Scaling code clone detection to big-code. In: Proceedings of the 38th International Conference on Software Engineering. Vienna. pp. 1157-1168.

[4] M. White, M. Tufano, C. Vendome and D. Poshyvanyk. (2016) Deep learning code fragments for code clone detection. In: Proceedings of the

31st IEEE/ACM international conference on automated software engineering. Singapore. pp. 87-98.

[5] Jiang, L., Misherghi, G., Su, Z., & Glondu, S. (2007) Deckard: Scalable and accurate tree-based detection of code clones. In: 29th International Conference on Software Engineering. Minneapolis. pp. 96-105.

[6] Haiduc, S., Aponte, J., & Marcus, A. (2010). Supporting program comprehension with source code summarization. In: Proceedings of the 32nd ACM/IEEE International Conference on Software Engineering-Volume 2. Cape Town. pp. 223-226.

[7] Jiang, S., Armaly, A., & McMillan, C. (2017) Automatically generating commit messages from diffs using neural machine translation. In: 2017 32nd IEEE/ACM International Conference on Automated Software Engineering (ASE). Urbana. pp. 135-146.

[8] Shi, E., Wang, Y., Du, L., Zhang, H., Han, S., Zhang, D., & Sun, H. (2021) Cast: Enhancing code summarization with hierarchical splitting and reconstruction of abstract syntax trees. arXiv preprint arXiv:2108.12987.

[9] Tairas, R., & Gray, J. (2009) An information retrieval process to aid in the analysis of code clones. Empirical Software Engineering, 14, 33-56.

[10] Liu, Y., Poshyvanyk, D., Ferenc, R., Gyimóthy, T., & Chrisochoides, N. (2009) Modeling class cohesion as mixtures of latent topics. In: 2009 IEEE International Conference on Software Maintenance. Edmonton. pp. 233-242.

[11] De Lucia, A., Di Penta, M., Oliveto, R., Panichella, A., & Panichella, S. (2012) Using IR methods for labeling source code artifacts: Is it worthwhile?. In: 2012 20th IEEE International Conference on Program Comprehension (ICPC). Passau. pp. 193-202.

[12] Harris, Z. S. (1954) Distributional structure. Word, 10(2-3), 146-162.

[13] Salton, G., & Buckley, C. (1988) Term-weighting approaches in automatic text retrieval. Information processing & management, 24(5), 513-523.

[14] Deerwester, S., Dumais, S. T., Furnas, G. W., Landauer, T. K., & Harshman, R. (1990) Indexing by latent semantic analysis. Journal of the American society for information science, 41(6), 391-407.

[15] Blei, D. M., Ng, A. Y., & Jordan, M. I. (2003) Latent dirichlet allocation. Journal of machine Learning research, 3(Jan), 993-1022.

[16] Eckart, C., & Young, G. (1936) The approximation of one matrix by another of lower rank. Psychometrika, 1(3), 211-218.

[17] Mikolov, T., Karafiát, M., Burget, L., Cernocký, J., & Khudanpur, S. (2010) Recurrent neural network based language model. In: Interspeech. Makuhari. Vol. 2, No. 3, pp. 1045-1048.

[18] Vaswani, A., Shazeer, N., Parmar, N., Uszkoreit, J., Jones, L., Gomez, A. N., ... & Polosukhin, I. (2017) Attention is all you need. In: 31st Conference on Neural Information Processing Systems. Long Beach. 30.

[19] Zhang, J., Wang, X., Zhang, H., Sun, H., Wang, K., & Liu, X. (2019) A novel neural source code representation based on abstract syntax tree. In: 2019 IEEE/ACM 41st International Conference on Software Engineering (ICSE). Montreal. pp. 783-794.

[20] Hua, W., & Liu, G. (2022) Transformer-based networks over tree structures for code classification. Applied Intelligence, 1-15.

[21] Bengio, Y., Simard, P., & Frasconi, P. (1994) Learning long-term dependencies with gradient descent is difficult. IEEE transactions on neural networks, 5(2), 157-166.

[22] Wang, Y. S., Lee, H. Y., & Chen, Y. N. (2019) Tree transformer: Integrating tree structures into self-attention. arXiv preprint arXiv:1909.06639.

[23] Child, R., Gray, S., Radford, A., & Sutskever, I. (2019) Generating long sequences with sparse transformers. arXiv preprint arXiv:1904.10509.

[24] Alon, U., Zilberstein, M., Levy, O., & Yahav, E. (2019) code2vec: Learning distributed representations of code. Proceedings of the ACM on Programming Languages, 3(POPL), 1-29.

[25] Baquero, J. F., Camargo, J. E., Restrepo-Calle, F., Aponte, J. H., & González, F. A. (2017) Predicting the programming language: Extracting knowledge from stack overflow posts. Advances in Computing: 12th Colombian Conference, CCC 2017, Cali, Colombia, September 19-22, 2017, Proceedings 12, pp. 199-210.

[26] Maletic, J. I., & Marcus, A. (2001) Supporting program comprehension using semantic and structural information. In: Proceedings of the 23rd International Conference on Software Engineering. Toronto. pp. 103-112.

[27] Ou, M., Cui, P., Pei, J., Zhang, Z., & Zhu, W. (2016) Asymmetric transitivity preserving graph embedding. In: Proceedings of the 22nd ACM SIGKDD international conference on Knowledge discovery and data mining. San Francisco. pp. 1105-1114.

[28] Tufano, M., Watson, C., Bavota, G., Di Penta, M., White, M., & Poshyvanyk, D. (2018) Deep learning similarities from different representations of source code. In: Proceedings of the 15th international conference on mining software repositories. Gothenburg. pp. 542-553.

[29] Ferrante, J., Ottenstein, K. J., & Warren, J. D. (1987) The program dependence graph and its use in optimization. ACM Transactions on Programming Languages and Systems (TOPLAS), 9(3), 319-349.

[30] Hu, X., Li, G., Xia, X., Lo, D., & Jin, Z. (2018) Deep code comment generation. In: Proceedings of the 26th conference on program comprehension. Gothenburg. pp. 200-210.

[31] Wei, H., & Li, M. (2017). Supervised deep features for software functional clone detection by exploiting lexical and syntactical information in source code. In: IJCAI. Melbourne. pp. 3034-3040.

[32] Dosovitskiy, A., Beyer, L., Kolesnikov, A., Weissenborn, D., Zhai, X., Unterthiner, T., ... & Houlsby, N. (2020) An image is worth 16x16 words: Transformers for image recognition at scale. arXiv preprint arXiv:2010.11929.

[33] Gulati, A., Qin, J., Chiu, C. C., Parmar, N., Zhang, Y., Yu, J., ... & Pang, R. (2020) Conformer: Convolution-augmented transformer for speech recognition. arXiv preprint arXiv:2005.08100.

[34] Hochreiter, S., & Schmidhuber, J. (1997) Long short-term memory. Neural computation, 9(8), 1735-1780.

[35] Ahmad, W. U., Chakraborty, S., Ray, B., & Chang, K. W. (2020) A transformer-based approach for source code summarization. arXiv preprint arXiv:2005.00653.

[36] Beltagy, I., Peters, M. E., & Cohan, A. (2020) Longformer: The long-document transformer. arXiv preprint arXiv:2004.05150.

[37] Zaheer, M., Guruganesh, G., Dubey, K. A., Ainslie, J., Alberti, C., Ontanon, S., ... & Ahmed, A. (2020) Big bird: Transformers for longer sequences. In: 34th Conference on Neural Information Processing Systems. Vancouver. 33, 17283-17297.

[38] Bahdanau, D., Cho, K., & Bengio, Y. (2014) Neural machine translation by jointly learning to align and translate. arXiv preprint arXiv:1409.0473.

[39] Rush, A. M., Chopra, S., & Weston, J. (2015) A neural attention model for abstractive sentence summarization. arXiv preprint arXiv:1509.00685.

[40] Brandes, N., Ofer, D., Peleg, Y., Rappoport, N., & Linial, M. (2022) ProteinBERT: a universal deep-learning model of protein sequence and function. Bioinformatics, 38(8), 2102-2110.

[41] He, K., Zhang, X., Ren, S., & Sun, J. (2015) Spatial pyramid pooling in deep convolutional networks for visual recognition. IEEE transactions on pattern analysis and machine intelligence, 37(9), 1904-1916.

[42] Mikolov, T., Chen, K., Corrado, G., & Dean, J. (2013) Efficient estimation of word representations in vector space. arXiv preprint arXiv:1301.3781.

[43] Puri, R., Kung, D. S., Janssen, G., Zhang, W., Domeniconi, G., Zolotov, V., ... & Reiss, F. (2021) Codenet: A large-scale ai for code dataset for learning a diversity of coding tasks. arXiv preprint arXiv:2105.12655.

[44] Deng, J., Dong, W., Socher, R., Li, L. J., Li, K., & Fei-Fei, L. (2009) Imagenet: A large-scale hierarchical image database. In: 2009 IEEE conference on computer vision and pattern recognition. Miami. pp. 248-255.

[45] Kitaev, N., Kaiser, Ł., & Levskaya, A. (2020) Reformer: The efficient transformer. arXiv preprint arXiv:2001.04451.

[46] Wang, S., Li, B. Z., Khabsa, M., Fang, H., & Ma, H. (2020) Linformer: Self-attention with linear complexity. arXiv preprint arXiv:2006.04768.

[47] Kingma, D. P., & Ba, J. (2014) Adam: A method for stochastic optimization. arXiv preprint arXiv:1412.6980.

[48] Sun, F., Liu, J., Wu, J., Pei, C., Lin, X., Ou, W., & Jiang, P. (2019) BERT4Rec: Sequential recommendation with bidirectional encoder

representations from transformer. In: Proceedings of the 28th ACM international conference on information and knowledge management. Beijing. pp. 1441-1450.

APPENDIX

This section selects representative categories from the algorithmic knowledge point collection used by SACC, and briefly introduces them with code examples from the dataset.

*A. Brute Force*

Brute Force is a simple and direct algorithm. It exhaustively enumerates all possible input combinations, maps each input to the corresponding output and records it. In the actual solving process, the result is directly queried to obtain the solution to the problem. The following code snippet, which prints a multiplication table, is a typical example of a Brute Force algorithm.

TABLE I. BRUTE FORCE

**Algorithm 1 Brute Force**

```
int main(){
  int i, j;
  for(i = 1; i < 10; i++){
    for(j = 1; j < 10; j++){
      printf("%dx%d=%d\n", i, j, i * j);
    }
  }
  return 0;
}
```

*B. Dynamic Programming*

Dynamic Programming is a commonly used algorithm, applicable to solving problems with overlapping sub-problems and optimal substructure characteristics. This method divides the original problem into a series of overlapping sub-problems, and uses the solutions of the sub-problems to construct the solution of the overall problem, thereby avoiding repeated calculations and improving algorithm efficiency. The key to the dynamic programming algorithm lies in determining the state transition equation, finding the relationship between states, and gradually deriving the optimal solution to the problem. The following code gradually calculates and fills the array to the target position according to the given pattern, that is, the value of $kai[i]$ is equal to the sum of the previous three positions.

TABLE II. DYNAMIC PROGRAMMING

**Algorithm 2 Dynamic Programming**

```
int dp(int n) {
  int kai[31];
  int i;
  kai[1] = 1; kai[2] = 2; kai[3] = 4;
  for (i = 4; i < 31; i ++) kai[i] = kai[i - 3] + kai[i - 2] + kai[i - 1];
  return kai[n];
}
```

*C. Sorting*

Sorting algorithms are used to arrange a set of data in an orderly manner according to certain rules, such as arranging the input data set in ascending or descending order, to facilitate faster operations such as retrieval and searching. Typical algorithms include bubble sort, quick sort, heap sort, and so on. This piece of code calculates the sum of the elements at even index positions after the input set of integers is sorted in ascending order.

TABLE III. SORTING

**Algorithm 3 Sorting**

```
int cmpint(const void *x, const void *y) {
  return *(int *)x - *(int *)y;
}
int main() {
  int n, l[200];
  scanf("%d", &n);
  for (int i = 0; i < 2 * n; i++) {
    scanf("%d", &l[i]);
  }
  qsort(l, 2 * n, sizeof(int), cmpint);
  int ans = 0;
  for (int i = 0; i < n; i++) {
    ans += l[2 * i];
  }
  printf("%d\n", ans);
  return 0;
}
```

*D. Arithmetic*

Arithmetic algorithms pertain to mathematical computations, performing basic arithmetic operations such as addition, subtraction, multiplication, division, high-precision operations, and base conversions. These algorithms primarily deal with numerical computations and processing. The following code iteratively reads two numbers input by the user and calculates the number of digits after addition.

TABLE IV. ARITHMETIC

**Algorithm 4 Arithmetic**

```
int main(){
  int i, n, r, n1, j, n2, count = 0;
  while(scanf("%d %d", &n1, &n2) == 2){
    n = n1 + n2;
    count = 0;
    while(n != 0){
      r = n % 10;
      count++;
      n = n / 10;
    }
    printf("%d\n", count);
  }
  return 0;
}
```

*E. Graph Theory*

Graph theory algorithms are a category of algorithms used to solve problems related to graph structures. Commonly used algorithms include Breadth-First Search (BFS), Depth-First Search (DFS), shortest path, minimum spanning tree, etc. These algorithms can be used to analyze and process various types of graphs, including directed and undirected graphs. The following code uses DFS to calculate the number of paths in an undirected graph that start from vertex 0, pass through all other vertices exactly once, and then return to vertex 0.

TABLE V. GRAPH THEORY

**Algorithm 5 Graph Theory**

```
int l[8][8];
int n, m, i, j, k, c;
void dfs(int s,int d,int M[]){
  int L[8], x;
  for(x = 0; x < n; x++) L[x] = M[x];
  for(x = 0; x < n; x++){
    if(L[x] && l[s][x]){
      if(d != n - 1) {L[x] = 0;dfs(x, d + 1, L); L[x] = 1;}
      else c++;
    }
  }
}
int main(){
  int M[8];
  scanf("%d%d", &n, &m);
  for(i = 0; i < n; i++) for(j = 0; j < n; j++) {M[i] = 1; l[i][j] = 0;}
  M[0] = 0;
  for(i = 0; i < m; i++){
    scanf("%d%d", &j, &k);
    l[j - 1][k - 1] = l[k - 1][j - 1] = 1;
  }
  dfs(0, 1, M);
  printf("%d", c);
}
```

*F. Computational Geometry*

Computational geometry algorithms are a category of algorithms used to solve geometric problems. They involve operations and calculations on geometric objects such as points, lines, polygons, etc. In addition to calculating the distance and dot product between points and vectors, it also includes polygon area calculation, convex hull algorithms, etc. The following code calculates the change in coordinates based on the input angle and step length.

TABLE VI. COMPUTATIONAL GEOMETRY

**Algorithm 6 Computational Geometry**

```
int main(){
  double x, y;
  int a, b, r, i;
  x = 0;
  y = 0;
  r = 90;
  scanf("%d,%d", &a, &b);
  while(a != 0 || b != 0){
    x += a * cos((double)r / 180 * PI);
    y += a * sin((double)r / 180 * PI);
    r -= b;
    scanf("%d,%d", &a, &b);
  }
  printf("%d\n%d\n", (int)x, (int)y);
  return 0;
}
```

*G. String*

A string is a sequence composed of characters. String algorithms are used for string processing, which include matching, searching, sorting, etc. It is necessary to choose the appropriate algorithm according to the characteristics of the string and the specific problem. The following code implements the reversal of a string.

TABLE VII. STRING

**Algorithm 7 String**

```
int i, j;
char str[50];
int main(){
  scanf("%s", str);
  for(i = 0; i < strlen(str); i++) printf("%c", str[strlen(str) - i - 1]);
  printf("\n");
  return 0;
}
```